# Nanoscale Mapping of Strain-engineered Excitons in Monolayer $WSe_2$

*Anna S. Roche[1], Nicholas V. Proscia[3], Paul D. Cunningham[3], Kathleen M. McCreary[3], Dante J. O'Hara[3], Hsun Jen Chuang[3], Cory Cress[3], Collin Sanborn[3,4], Samuel W. LaGasse[4], Jose J. Fonseca[3], Jeremy T. Robinson[3], Takashi Taniguchi[5], Kenji Watanabe[6], John R. Schaibley[1,2], Brian J. LeRoy[1*]*

[1]*Department of Physics, University of Arizona, Tucson, Arizona 85721, USA*
[2]*Wyant College of Optical Sciences, University of Arizona, Tucson, Arizona 85721, USA*
[3]*United States Naval Research Laboratory, Washington, District of Columbia 20375, USA*
[4]*Laboratory for Physical Sciences, 8050 Greenmead Drive, College Park, MD, 20740 USA*
[5]*Research Center for Materials Nanoarchitectonics, National Institute for Materials Science, 1-1 Namiki, Tsukuba 305-0044, Japan*
[6]*Research Center for Electronic and Optical Materials, National Institute for Materials Science, 1-1 Namiki, Tsukuba 305-0044, Japan*
**Corresponding Author*

Excitons in monolayer transition metal dichalcogenides (TMDs) are highly sensitive to external stimuli, allowing control of resonance energies and lifetimes through electric and magnetic fields, heterostructures, moiré potentials and mechanical strain. Observations of single-photon emission (SPE) near sample edges, wrinkles, and strained nanobubbles have motivated strain-engineering aimed at realizing localized emitters in two-dimensional materials. The origin of these emitters remains unclear and prior studies have relied on far-field spectroscopy with strain profiles inferred separately through topography, limiting direct correlation between local strain and excitonic response. Here, we directly probe strain-modified excitonic spectra in monolayer $WSe_2$ using cryogenic scattering-type scanning near-field optical microscopy (s-SNOM). We compare two strain-engineering techniques, nano-indentation and patterned substrates used to produce SPEs. At 9 K, we spatially resolve exciton strain-induced resonance shifts with sub-diffraction limit resolution by simultaneously measuring near-field optical response and topography. We observe that strain transferred from nano-indentation into a deformable polymer substrate differs from that produced by draping TMDs over pre-patterned dielectric pillars. Adhesion and tearing during plastic deformation drive a compressive exciton blueshift, while draping produces dominant tensile strain. Comparing the strain model results to the experimentally observed energy shifts shows that the nano-indentation strain is over an order of magnitude smaller than previously inferred, due to puncturing of the fragile monolayer, whereas pillar-strained devices remain intact under more modest strain. These differences suggest SPE arises from complex, strain-mediated interactions rather than simple confinement potentials, establishing s-SNOM as a powerful probe of quantum emission in strained 2-dimensional materials.

## 1. INTRODUCTION

Monolayer transition metal dichalcogenide (TMD) semiconductors have emerged as a model platform for exploring the consequences of reduced dimensionality. In the monolayer limit, reduced dielectric screening enhances Coulomb interactions, strengthening electron–hole attraction and stabilizing tightly bound excitons. The large exciton binding energy dominates the optical response across the visible and near-infrared spectral range owing to the strong oscillator strength and efficient coupling to light[1–5]. The same reduced dimensionality that amplifies these Coulomb correlations also makes excitonic properties sensitive to external perturbations. Mechanical strain directly modifies the local band structure, enabling spatial control of exciton resonance energies and confinement on nanometer length scales. Reports of single photon emission (SPE) in TMD materials[6–8] near sample edges[9] and naturally occurring, highly strained bubbles[10] and wrinkles inspired various strain engineering studies[11,12] towards realizing deterministic SPEs in 2D materials. The recent realization of SPEs in 2D materials is particularly compelling for the development of placeable, on-chip single-photon emitters, since thin monolayer materials can be easily and controllably strained. Such highly strained regions around bubbles, wrinkles, and edges raise the question of the role nanoscale defects and disorder play in the observed emission[13–15].

Previous measurements of SPEs in highly strained TMDs[15] have primarily relied on far-field optical spectroscopy techniques combined with strain profiles extracted from separate AFM measurements. Such analysis reported strain percentages as high as -2 to 4 percent within a selected 200 nm region, which would correspond to A exciton energy shifts of several hundred meV[16,17]. However, to observe the behavior in these highly strained areas requires nanometer scale spatial resolution. The foundational room-temperature tip-enhanced emission measurements show red shifted emission at the base of bubbles in $WSe_2$ on the order of tens of meV[18], and subsequent studies see similar behavior in wrinkles and pillars[19–21], but emission is sensitive both to excitons and to emissive defect states. Exciton absorption resonances offer an unambiguous probe of local strain, yet no near-field study has directly examined how they shift under strain. Cryogenic visible s-SNOM (Figure 1A) has demonstrated[22] the ability to detect energy shifts below 1 meV through exciton resonances, ensuring sufficient sensitivity for precise strain measurements.

In this work, we directly probe the nanoscale excitonic response, first, of AFM-indented[23] (Figures 1,2,3) and then pillar strained[19] (Figure 5) monolayer $WSe_2$ at cryogenic temperatures using resonant visible (~710 nm) scattering-type scanning near-field optical microscopy (s-SNOM). Nanoscale optical measurements were performed with a cryogenic s-SNOM (cryo-neaSCOPE), operational from room temperature to 9 K, equipped with a tunable, narrow linewidth continuous-wave laser. S-SNOM[24–29] builds on atomic force microscopy (AFM) principles: a metal-coated AFM tip is illuminated by a focused laser, acting as a nanoscale antenna that enhances the optical field at the tip apex (Figure 1A). This field confinement enables sub-diffraction optical resolution (~40 nm), roughly ten times tighter than the diffraction limit at the low-temperature $WSe_2$ exciton resonance (~710 nm). When brought into proximity with the sample, near-field interactions modify the light backscattered from the tip. The scattered signal is collected in the far field and analyzed via pseudo-heterodyne interferometric detection (PHID)[30], which demodulates the signal at higher harmonics of the tip oscillation frequency ($\Omega$) to isolate the near-field contribution from the dominant far-field background (Figure S1). Throughout this work, detection is performed at either the fourth or fifth harmonic, with the choice determined by the signal-to-noise ratio (SNR). The complex scattering coefficient, $\sigma = se^{i\phi}$, encodes the amplitude ($s$) and phase ($\phi$) of the near-field signal, providing local information on the dielectric function and excitonic resonances. Raster scanning the tip across the sample allows simultaneous mapping of topography, near-field amplitude ($s$), and phase ($\phi$) as shown in Figure 1D and 1E.

## 2. RESULTS AND DISCUSSION

We first study the exciton resonance in samples strained by AFM-indentation technique. The $WSe_2$ monolayers are wet-transferred onto a poly(methyl methacrylate) (PMMA) coated substrate, then indented (Figure S2) following a 7 by 7 grid pattern depicted in Figure 1B. Additional growth and fabrication details can be found in the Methods section. To investigate the exciton behavior as a function of strain, we spatially map the A exciton resonance as a function of excitation energy. We repeated line scans along a column of the $WSe_2$ nano-indentations, marked in Figure 1B, while stepping the excitation energy through the A exciton absorption resonance near 1.75 eV in increments of ~0.7 meV. The amplitude and phase at the fourth and fifth harmonic of the tip oscillation frequency were normalized[25] against the substrate (PMMA – nano-indentation and hBN - pillar) signal as $s_5(\boldsymbol{r})/s_{5,sub}$ and phase $\phi_5(\boldsymbol{r}) - \phi_{5,sub}$ as neither PMMA or hBN have an energy dependent optical response in the energy range of the A resonance. The substantial height variation from AFM indentation of the TMD can increase background artifacts, necessitating higher-harmonic tip demodulation. The region on the right of Figure 2A and B shows the included PMMA region necessary for proper normalization.

The indented regions, displaying negative Z values across the height profile in Figure 2C (left axis), appear as the black vertical stripes in near-field amplitude and the red regions in the near-field phase in Figure 2A and 2B respectively. These features appear from the absence of a resolvable exciton resonance within the measured energy range at those spatial locations. A parabolic feature is observed between each of the indents in both amplitude and phase, as the exciton resonance appears to blue shift progressively when approaching each indent edge. Labelled indents six and seven straddle a region of trilayer $WSe_2$ and the region to the right of indent seven is majority bilayer, with an optical response outside of the probed energy range. Repeated measurements on bare PMMA do not exhibit any resonance within the probed energy, confirming the response measured is from $WSe_2$, and not from unwanted background artifacts along a highly deformed substrate (Figure S3).

To quantify the observed energy shift, the near-field amplitude and phase spectra were extracted at each pixel along the line scan. Each spectrum was fit using the well-established Finite Dipole Model (FDM)[31,32] (Supporting Information Section 4). The extracted resonance center energy at each pixel is shown in Figure 2C, with the right-hand axis scaled to match the probed energy range shown in Figure 2 A-B. The left axis shows the simultaneously measured height profile. When the height profile is overlaid with the extracted A exciton center energy, the exciton energy is observed to repeatably blue shift when approaching the rim of pushed out material around each indent, in good agreement with far-field reflectivity mapping of similar samples (Figure S4). Focusing on the exciton energy behavior along the nano-indentation sites (Figure 2D), the exciton energy shifts from 1.746 eV at the unstrained midpoint (Figure S5) between two indents, denoted with a solid red line, to 1.76 eV nearing the indentation rim (green solid line). This ~15 meV blue shift occurs with an out-of-plane height deformation of roughly 10 nm, as illustrated by the overlaid height profile in Figure 2D. This smooth blue shift is consistently observed across the measured indents, suggesting compressive rather than tensile strain dominates in such regions. Further up and along the indentation edge, where the PMMA height rises to roughly 60 nm before falling into the indent, an energy could not be extracted, as no clear exciton resonance was present within the probed spectral range.

The measured spectra align with our understanding of the nano-indentation process. The transferred TMD is adhered to the deformable polymer substrate. As the AFM tip indents the polymer, it undergoes plastic flow[23] as material from inside the crater is pushed up and out around the tip, piling up to create the observable raised rim. Large tensile strain is expected as the adhered TMD stretches along the steep indent sidewalls. At the outer rim, however, the polymer is displaced laterally, shortening the surface and compressing the adhered monolayer.

Prior measurements of biaxially strained $WSe_2$ at 10K report A exciton energy shifts around -63 to -120 meV/% strain[16,17]. Following previous works[15], modelling our sample as a monolayer adhered to the deformed PMMA layer experiencing biaxial strain (Supporting Information Section 7) predicts unphysical strains ranging from -10 to 40% (Figure S6), well beyond the breaking strain of monolayer TMDs[33]. Strain calculated from the post-indentation topography does not account for the plastic deformation the PMMA and adhered monolayer undergo during indentation. Our direct near-field measurements reveal local blueshifts of only ~15 meV, indicating that the local strain is compressive and suggesting the true strain transfer is far below the often assumed 100%.

We attribute the lack of resolvable absorption peaks near the edge of the indentation crater, where the s-SNOM signal drops off sharply, to material tearing. Rather than adhering perfectly across the deformed indent, the $WSe_2$ monolayer likely tears within the deep crater to relieve the extreme tensile strain. This is consistent with tears visible in high-resolution AFM images (Figure S7) and recent SEM imaging[34], the abrupt disappearance of the exciton resonance at the indentation edges, and with far-field maps (Figure S4) taken over a larger energy window. These results show that strain estimates based on height topography alone, which assume the 2D layer is structurally continuous and that substrate strain transfers perfectly, drastically overestimate the true strain in nano-indented 2D materials and cannot capture the contributions from plastic flow; our near-field spectral maps instead capture the reduced, tear-limited strain actually present in these samples.

The measurements in Figure 2 were taken along a single line-scan across the sample. Based on the compressive strain trend discussed above, this behavior is expected to hold along the entire rim, motivating a full near-field-resolved map of the strain distribution around the indent. Mapping was performed on a second indented $WSe_2$ sample (Figure 1D, 1E, S8) by repeatedly imaging an 8.5 by 3 µm region centered on a representative indent while stepping the excitation energy from 1.73 to 1.76 eV in 1 meV increments, thereby measuring the near-field spectra at each pixel. Representative near-field images at 1.748 eV are shown in Figure 3B, C. Following the same FDM fitting procedure, we extract the complex dielectric function from each near-field spectrum. Figure 3D shows the extracted A exciton center energy, $\omega_0$, across the $WSe_2$ region. The center energy of the exciton resonance shows a ring-like blue shift in the bottom left region of the image, moving towards the indentation site, in good agreement with the energy shifts measured by earlier line cuts (Figure 2).

This feature indicates the surface displacement of the rim is likely the dominant source of compressive strain as the blue shift maps to the rim formed around each nano-indentation crater. This rim forms from plastic flow of the substrate during the indentation process (Figure 4A, 4B). As the rim volume is laterally displaced (Figure 4C) it compresses the adhered $WSe_2$, while the monolayer within the crater experiences extreme tensile strain as it is pushed into the PMMA by the AFM tip.

It is of interest to compare the strain profile produced by AFM indentation with the more common strain-engineering approach of draping a TMD over a dielectric pillar. The draped TMD is expected to experience tensile strain (Figure 4E, 4F) as it tents over the structure, the combination of substrate adhesion and thin-film bending causes buckling[35], Figure 4G, causes wrinkles to develop between pillars. Near the wrinkles, the profile is a mix of compressive and tensile strain[21], Figure 4H. Figure 5 shows an hBN-capped $WSe_2$ monolayer transferred onto 60-80 nm-high fabricated $SiO_2$ pillars (Methods). The line-scan measurements presented in Figures 5C, D were acquired along row five, as indicated in Figure 5A. These line scans show the strain induced center energy shifts of the A exciton measured by the local amplitude (Figure 5C) and phase (Figure 5D) of the near-field spectra. Figure 5E displays the topographic height profile together with the FDM-derived A exciton center energy evaluated at each pixel along the scan. Compared to the AFM indentation sample, the energy shifts induced by the pillars are less symmetric and show both blue and red shifts. This trend agrees with previous studies of pillar-strained TMDs, which show that the monolayer develops wrinkles and asymmetric strain distributions during relaxation[19,35].

The strain modelling was repeated on a single pillar, shown in Figure 5B. Topographic mapping suggests low tensile strain <0.6% in the area surrounding the pillar, with a sharp increase to ~1% directly at the pillar contact interface as seen in Figure 5F. From the calculated strain profile, while approaching the pillar, the A exciton energy is expected to shift from roughly 1.75 to 1.71 eV before sharply jumping to 1.68 eV on the pillar surface, in agreement with previous far field results on a comparable system[36]. Using the data in Figure 5C, D, we directly measure energy shifts from 1.76 to 1.725 eV as shown by the markers in Figure 5G. The modelled results are qualitatively consistent with several of our measured pillars, differences we attribute to the TMD relaxing non-uniformly over a large, gridded area rather than around a single isolated pillar. In Figure 5C, D, there is a clear red shift around the pillars at 11 and 30 microns showing the tensile strain as the TMD deforms around the raised pillars. These measurements were repeated along Row 4, and a similar energy profile is seen around the pillar at 25 microns in Figure S9.

The strain variation between nano-indents and pillars suggests that strain-induced confinement alone may not explain the low temperature single photon emission common to both techniques (Figures S10, S11). It may be that single photon emission emerges from the more complex locally varying nanoscale strain around wrinkles and tearing introduced in the process. Rather than acting as a drawback, tears may provide a high density of edge-states and dangling bonds to act as emissive defect sites. This has been recently investigated by nano-indenting TMDs on a less deformable surface[37], to purposefully tear and fold the fragile TMD, resulting in SPE emission from ~76% of nano-indents.

## 3. CONCLUSIONS

This work establishes cryogenic visible s-SNOM as a robust, local, and non-destructive technique for measuring strain. We present previously unattainable measurements of local absorption shifts induced by common TMD strain-engineering techniques. Our measurements show that nano-indentation yields a significantly different strain environment than draping TMDs over pillars, due primarily to adhesion to the deformable polymer under plastic flow. These results suggest that AFM nano-indentation likely punctures the fragile monolayer, leading to over an order-of-magnitude reduction in the achieved strain compared to estimates based on the gradient of the topography. The fact that both strain engineering techniques produce high yields of SPEs suggests these emitters are not simply strain confined excitons and instead emerge due to the complex local environment which warrants further study. The observed tears from nano-indentation generate edge defects and dangling bonds that introduce sub-bandgap defect states, which must be considered in explaining SPE formation[15]. Future work correlating SPE location with strain maps and shifts to the absorption bands may be a promising path toward understanding the nanoscale mechanisms that activate these emitters. Beyond SPE device applications, cryogenic nano-imaging enables direct measurement of local strain through shifts in resonance energy. As s-SNOM is operational from the visible to the terahertz, this approach is readily extendable across a wide range of materials systems allowing direct measurement of total percent strain transferred to a material in future strain engineering devices.

## 4. METHODS

### 4.1 Chemical vapor deposition TMD growth recipe

CVD grown TMD samples were used for the AFM indentation sample. $WSe_2$ monolayers were grown on 275nm $SiO_2$/Si substrates via chemical vapor deposition (CVD) at ambient pressure in a two-inch quartz tube furnace. Perylene-3,4,9,10-tetracarboxylic acid tetrapotassium salt (1mM) was drop-cast onto the substrates to promote lateral $WSe_2$ growth. Solid $WO_3$ (Alfa Aesar, Product No. 13398) and selenium (Alfa Aesar, Product No. 36208) were used as precursors, with a continuous argon flow (65 sccm) during the temperature ramp to 800°C. Upon reaching 800°C, hydrogen (10 sccm) was introduced and maintained during the 10-minute soak and subsequent cooling.

4.2 AFM Nano-indentation

Strained TMD monolayers shown in Figures 1-3 were fabricated using AFM tip indentation[23]. For Sample 1, CVD grown TMD monolayers are wet-transferred to 300 nm thick PMMA spin-coated onto a Distributed Bragg Reflector (DBR) with a 750 nm center wavelength and a ~220 nm stop band. The DBR consisted of 10.5 periods of alternating $TiO_2$ and $SiO_2$ layers with nominal thicknesses of 78.57 nm and 125.85 nm, respectively and terminated with a top $SiO_2$ layer. The DBR was fabricated on a silicon substrate via sputtering. The DBR structure does not modify the near-field results in this study but was fabricated to assist with room-temperature far-field measurements of interest to collaborators. Sample 2 measured in Figure 3 was fabricated on a PMMA coated 295 nm $SiO_2$/Si substrate, with an optical image shown in Figure S8.

The TMD and underlying deformable PMMA layer are indented with a standard non-contact mode AFM tip (26 N/m force constant) using the spectroscopy toolset on a Park XE AFM. The indentation depth can be varied to produce different deformation profiles. The indentations presented in the main text were created with a Z-depth setpoint of 2.75 microns, which produces an indentation depth of roughly 150 nm at the base of the indent. Additional images and details in Figure S2.

4.3 Pillar substrates

The pillar substrate was fabricated from a 280 nm $SiO_2$/Si wafer. Negative Ma-N 2403 resist was spin coated and then exposed in a Raith Voyager electron beam lithography (EBL) system. The EBL pattern exposed 50-100 nm circles of the resist on the site of each pillar, following the 5 by 5 micron grid pattern shown in Figure 5. The resist was then developed with MF-319. Approximately 80 nm tall pillars were formed by etching the underlying $SiO_2$ in $CHF_3$/Ar in an Oxford PlasmaPro reactive ion etcher, using the remaining resist 'mesas' as the mask. After etching, the resist is fully removed with Remover PG and O2 plasma ashing.

The pillar sample was assembled with gold template[38] exfoliated monolayer $WSe_2$ and mechanically exfoliated hBN. The hBN/$WSe_2$ structure was assembled by using a polycarbonate (PC) based dry-transfer method[39], first picking up the hBN and then the monolayer $WSe_2$. The van der Waals stack is then melted onto the prefabricated pillar substrate at 165 °C. Finally, the PC is removed with dichloromethane.

4.4 Optical measurements

All optical measurements were performed in a cryo-neaSCOPE using a Pt/Ir-coated silicon AFM tip (PPP-NCLPt). The cryogenic s-SNOM is held in an optical cryostat at 9 K. Near-field amplitude and phase images were measured with a tunable visible Ti-Sapphire laser (MSquared SolsTiS), fiber coupled into the s-SNOM optical system and subsequently power (4 mW) and wavelength locked to ensure proper stability. All near field images were detected at the 4$^{th}$ (Figure 3) and 5$^{th}$ (Figures 2,5) harmonic of the AFM tapping frequency to maximize signal-to-noise ratio and ensure good near-field confinement.

4.5 Finite Dipole Modeling

The dielectric response was calculated from the Finite Dipole Model (FDM), implemented via an open-source Python package snompy[40]. To capture the layered geometry of the sample, the sample was modeled as a multilayer stack following the transfer matrix method[31,41]. Additional details and variable parameters in Supporting Information.

**Corresponding Author**

* Brian LeRoy - Department of Physics, University of Arizona, Tucson, Arizona 85721, USA. leroy@arizona.edu

**Author Contributions**

ASR, BJL, NVP, and PDC conceived the project. BJL and JRS supervised the project. ASR performed the experiments and analyzed the data with input from BJL, NVP, PDC, and JRS. HJC, DJO, and ASR fabricated the nano-indentation samples. KMM grew and characterized the CVD grown TMD crystals and performed far-field characterization. SWL, CS, CC, JJF, and JTR fabricated and characterized the nanopillar TMD devices. NVP performed SPE characterization. TT and KW provided bulk hBN crystals. ASR wrote the paper with input from BJL, PDC, and JRS. All authors discussed the manuscript.

**Notes**

The authors declare no competing financial interest.

**Funding Sources**

The work at the University of Arizona was supported by the National Science Foundation under grants DMR-1828427, DMR-2003583 and ECCS-2122462 and AFOSR Grant Nos. FA9550-22-1-0113, FA9550-22-1-0220 and FA9550-22-1-0312 and by the Army Research Office under grant W911NF-20-1-0215. The work at NRL was supported by base programs and the NRL Nanoscience institute. KW and TT acknowledge support from the JSPS KAKENHI (Grant Numbers 21H05233 and 23H02052), the CREST (JPMJCR24A5), JST and World Premier International Research Center Initiative (WPI), MEXT, Japan.

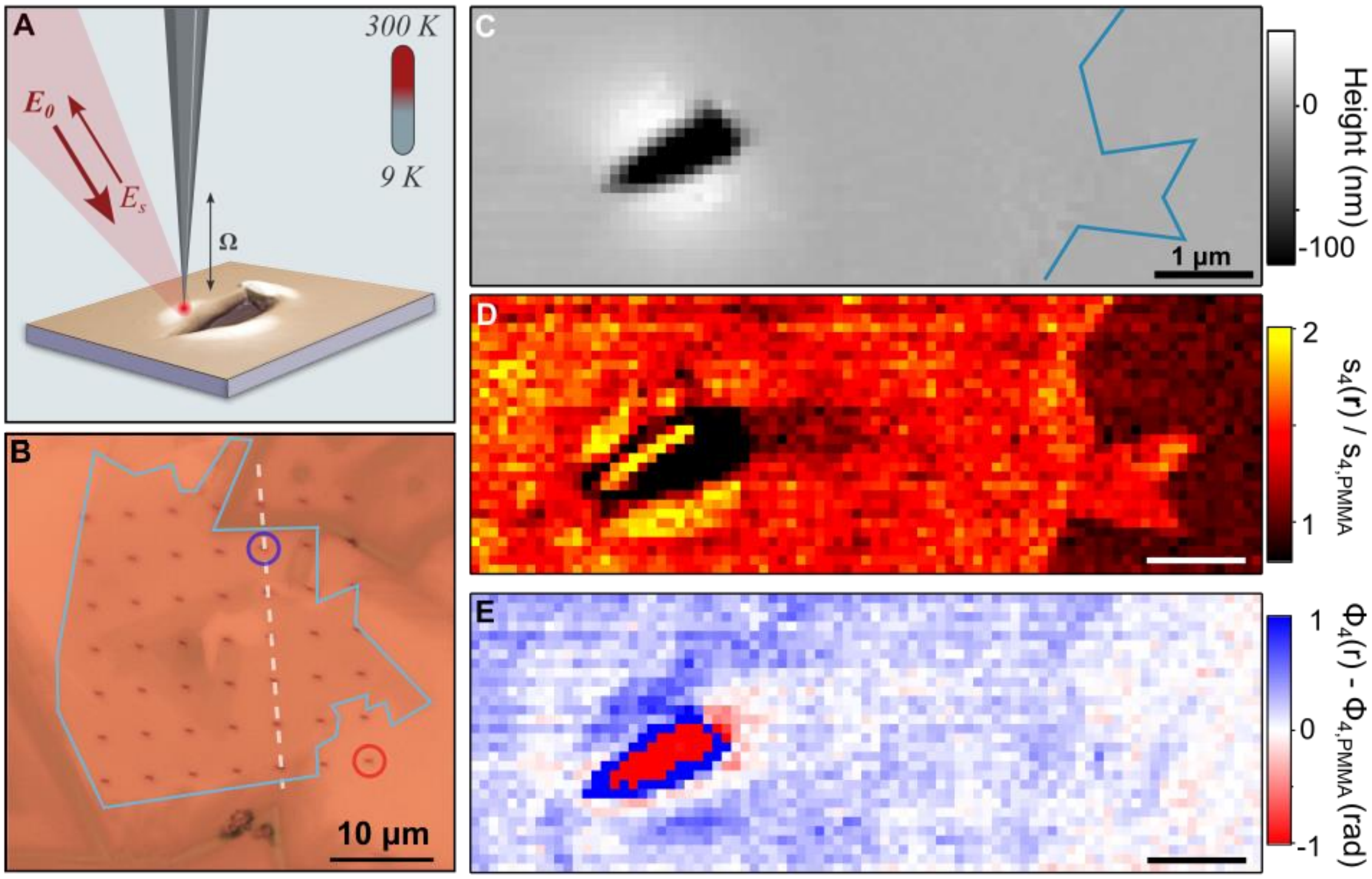

**Figure 1.** Near-field measurements of AFM indented $WSe_2$. **A** Schematic of the s-SNOM experimental set-up. A tunable visible laser is focused on a metallic AFM tip, operating in tapping mode at frequency Ω, to produce a nanoscale enhanced spot at the tip apex. The near-field spot interacts with the sample below the tip and backscatters to the focusing optic to be sent to a far field detector. The microscope is installed on a cold head and held in an optical cryostat at 9 K. **B** An optical microscope image of Sample 1 depicting the grid of AFM indentation points across a region of monolayer (blue outline) $WSe_2$. The white line marks the s-SNOM line-scan path measured in Figure 2. The blue circled region is used for strain modelling, while the red circled region is used as bare PMMA control (Figure S3). Scale bar: 10 microns. **C** AFM topography image of nano-indented monolayer (blue outline) $WSe_2$ from Sample 2. Scale bar: 1 micron. **D** Normalized near-field amplitude ($s_4$) image of the sample taken at the 4th harmonic of the tapping frequency taken at an excitation energy of 1.73 eV across the region shown in **C.** **E** The corresponding normalized near-field phase ($\phi_4$) image of the sample.

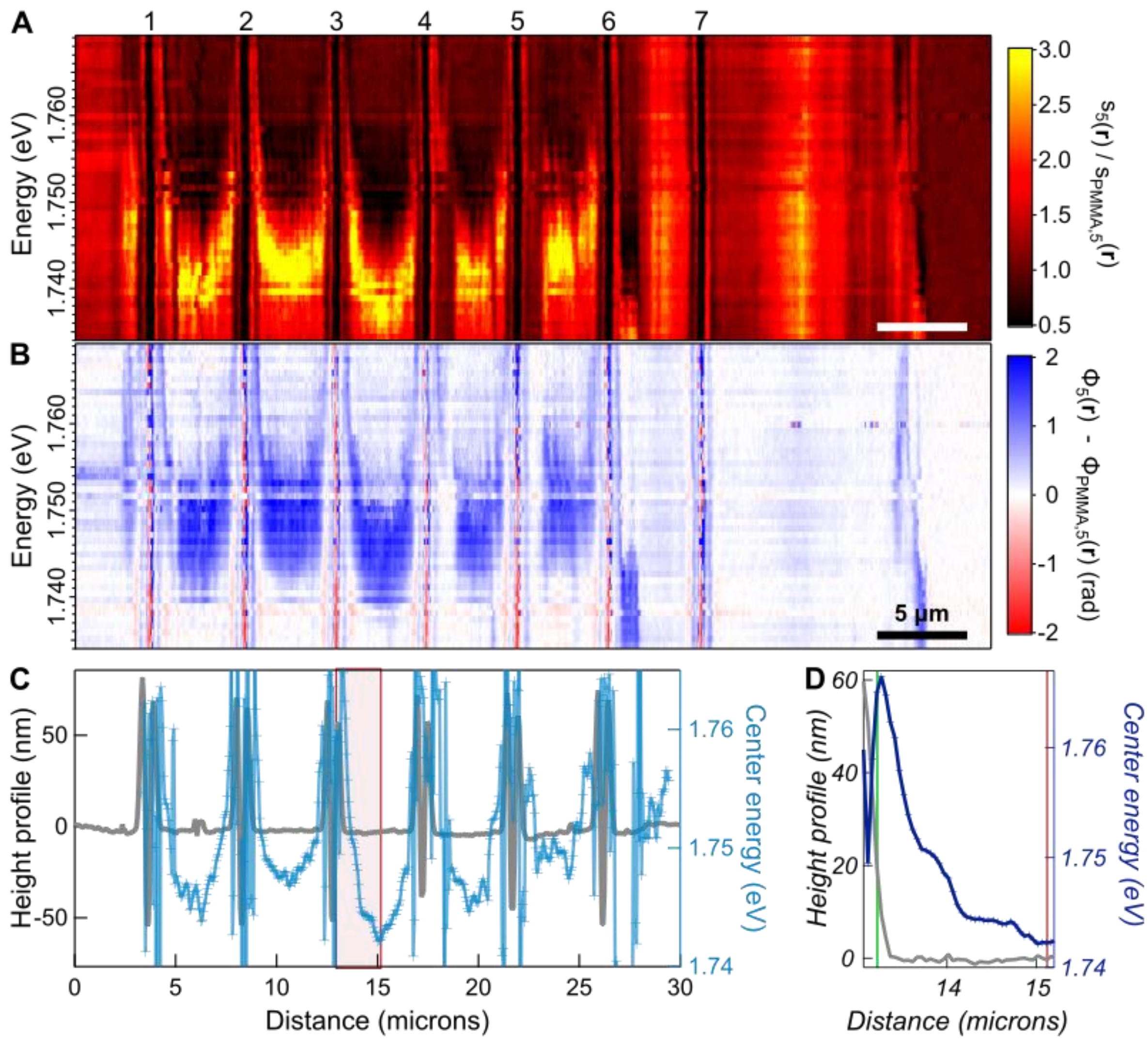

**Figure 2.** Exciton spectra measured across AFM indented $WSe_2$. **A,B** Variation of the near-field amplitude $s_5$ and phase $\phi_5$ signal across the line scan shown in Figure 1B as a function of excitation energy (left axis). Scale bars are 5 microns. **C** Sample height profile (black, left axis) along the scanned line in **A,B** with finite dipole model (FDM) extracted center energy of the A resonance extracted from a vertical line cut at each pixel (blue, right axis). **D** Representative exciton energy behavior moving from a strained region between two indents (green line, 13 microns) to an unstrained region between Indents 3 and 4 (red line, ~15 microns). Region of interest marked with red box in C.

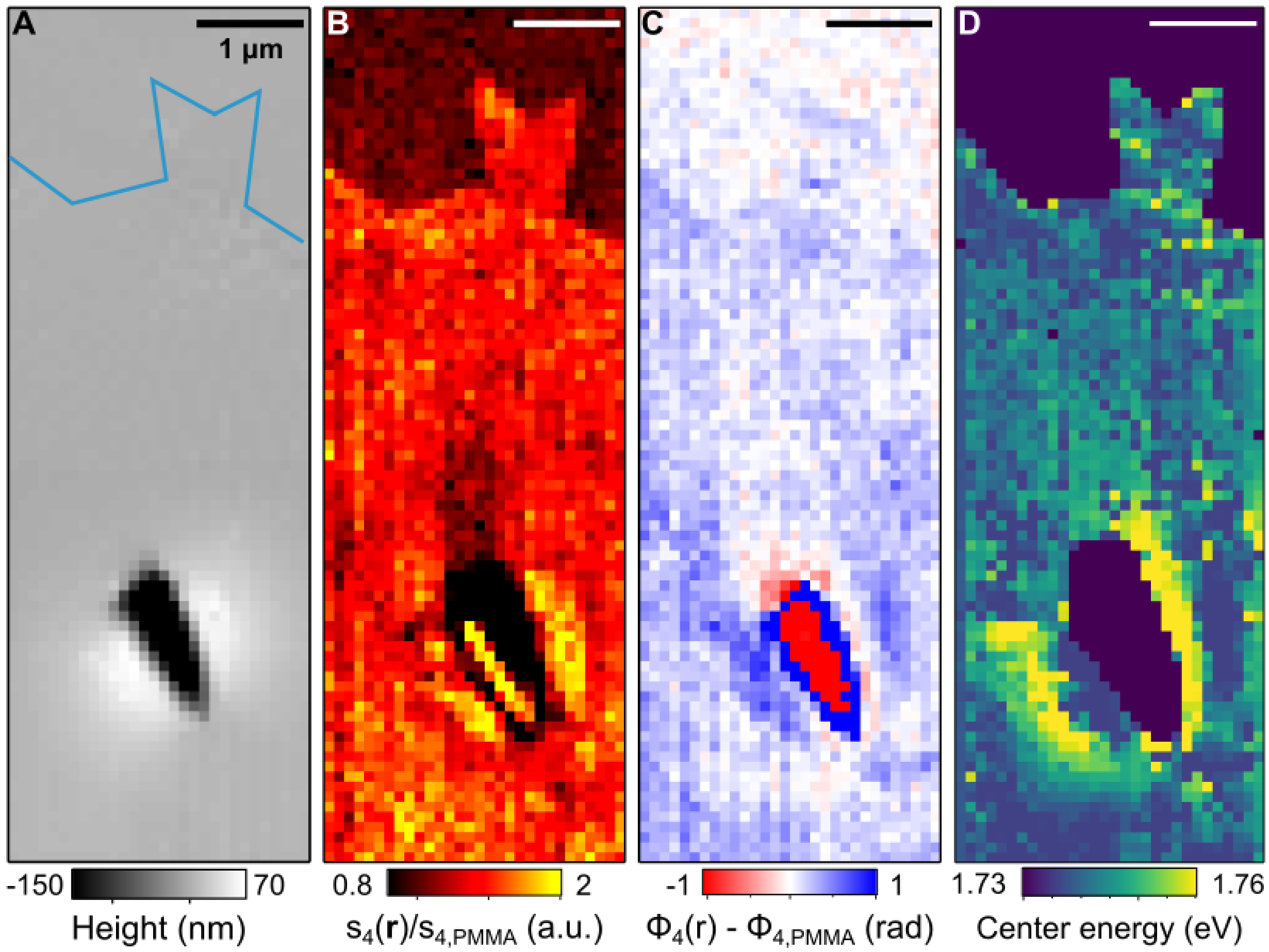

**Figure 3.** Mapping strain of nano-indentation region **A** Investigated region from Sample 2 (Figure S8) showing the amplitude (**B**) and phase (**C**) of the measured region at 1.748 eV, near the A exciton resonance. The same scan was repeated as the excitation energy was stepped through the A exciton resonance to map the A exciton center energy as a function of position around the indent region. **D** Resonance energy, $\omega_0$, value across the sample region extracted from the complex dielectric function at each pixel. Spatial disorder in the A resonance energy is observed across the $WSe_2$ flake, following the rim feature in **A**.

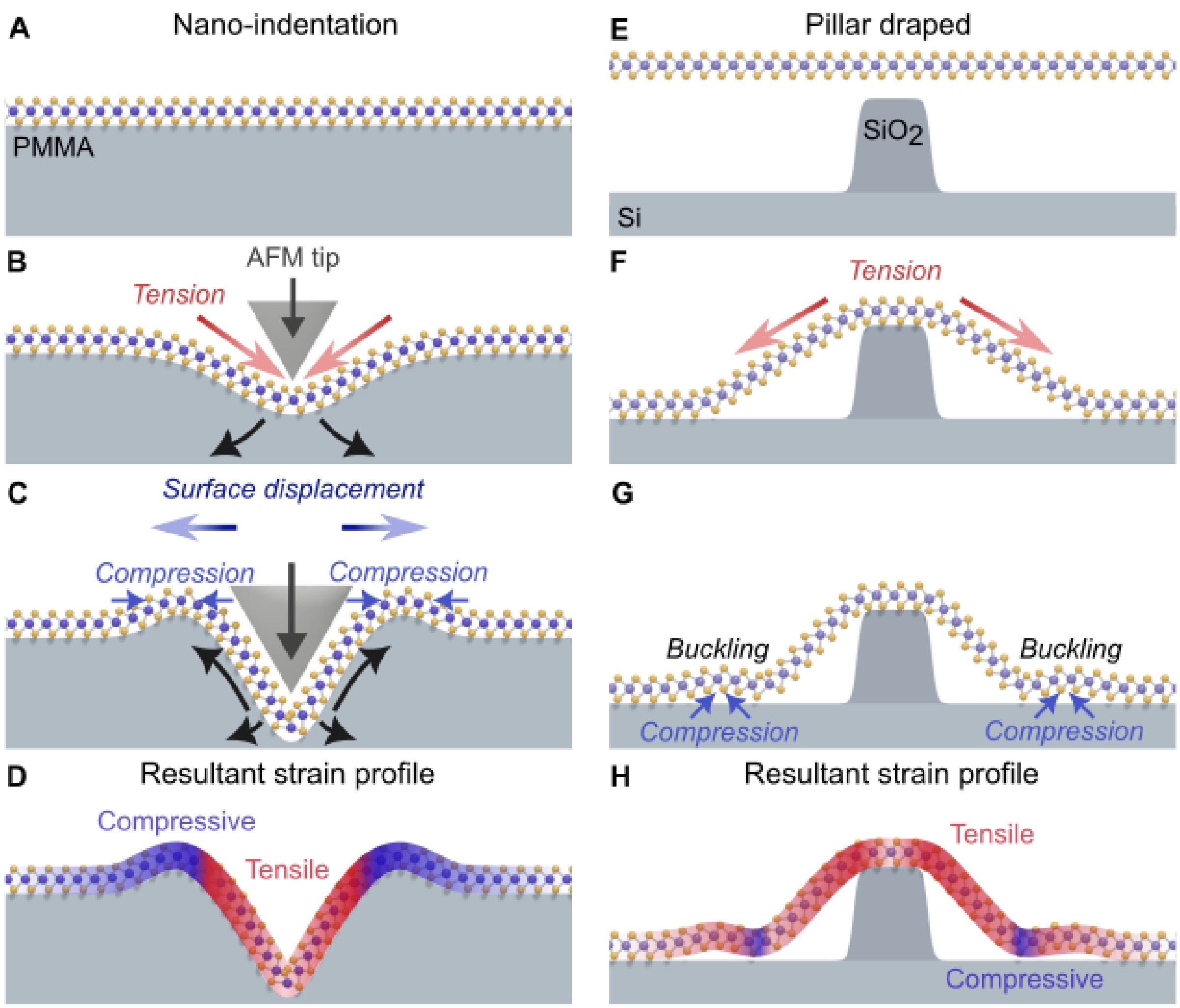

**Figure 4.** Scheme of nano-indentation and draping processes. **A** $WSe_2$ is transferred onto a PMMA substrate and an AFM tip creates the indentation, inducing tensile strain and initiating plastic flow of the deformable PMMA (**B**). The volume conservation during plastic flow creates a rim, which experiences lateral surface displacement causing compression of the adhered $WSe_2$ (**C**). The resulting strain profile in **D** is dominated by extreme tensile values, beyond breaking strain, in the crater and compression in the rim. **E** $WSe_2$ is transferred to a nanostructured substrate and draped over an individual dielectric pillar, leading to stretching and tensile strain (**F**). Because the flake must deform radially and across a gridded structure, the TMD forms wrinkles (**G**). The resulting strain in **H** is dominantly tensile with areas of alternating compressive and tensile strain near wrinkles.

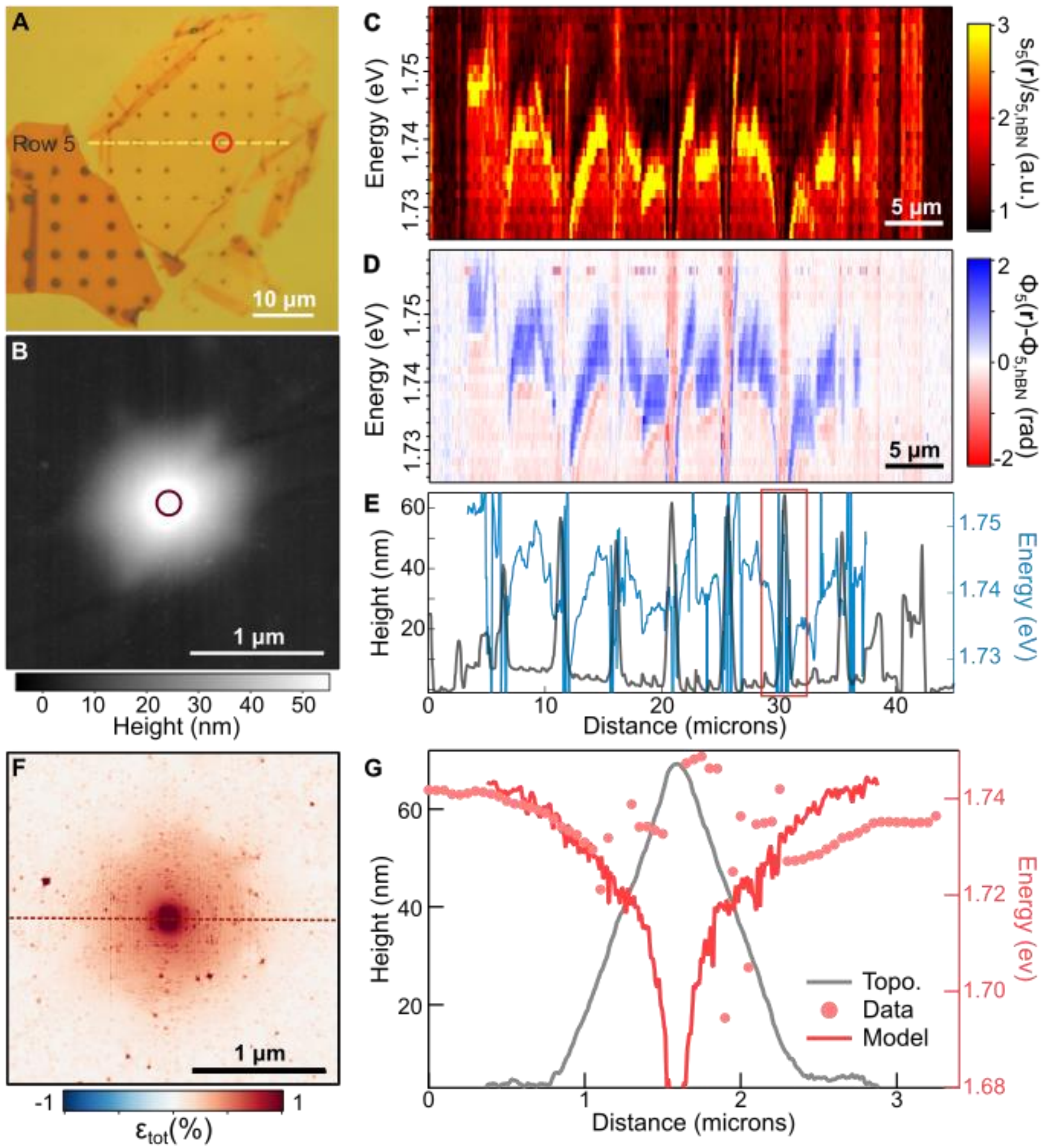

**Figure 5.** Exciton spectra of $WSe_2$ draped across $SiO_2$ pillars **A** Optical image of pillar strained sample showing deposited hBN capped $WSe_2$ on a grid of 60 nm tall cylindrical pillars with 5-micron spacing. **B** AFM topography image of the pillar circled in red which is representative of the sample deformation around each pillar. **C** Variation of near-field amplitude ($s_5$) across Row 5 shown in A as a function of excitation energy (left axis). **D** corresponding phase signal ($\phi_5$). Scale bars are 5 microns. **E** Sample height profile (gray, left axis) along the scanned line in **C,D** with finite dipole model (FDM) extracted center energy of the A exciton resonance extracted from a vertical line cut at each pixel (blue, right axis). Extracted center energy values shown for the region of clearly resolvable resonance behavior in **C**. **F** Strain percentage modelled from topography image in **B**. **G** The pillar height profile overlaid atop the expected energy shift of the A exciton extracted along the dashed line across modelled strain (**F**) compared to the near-field extracted energy shifts (pink markers).

**<u>Supporting Information:</u> Nanoscale Mapping of Strain-engineered Excitons in Monolayer $WSe_2$**

*Anna S. Roche[1†], Nicholas V. Proscia[3], Paul D. Cunningham[3], Kathleen M. McCreary[3], Dante J. O'Hara[3], Hsun Jen Chuang[3], Cory Cress[3], Collin Sanborn[3,4], Samuel W. LaGasse[4], Jose J. Fonseca[3], Jeremy T. Robinson[3], Takashi Taniguchi[5], Kenji Watanabe[6], John R. Schaibley[1,2], Brian J. LeRoy[1*]*

*[1]Department of Physics, University of Arizona, Tucson, Arizona 85721, USA*
*[2]Wyant College of Optical Sciences, University of Arizona, Tucson, Arizona 85721, USA*
*[3]United States Naval Research Laboratory, Washington, District of Columbia 20375, USA*
*[4]Laboratory for Physical Sciences, 8050 Greenmead Drive, College Park, MD, 20740 USA*
*[5]Research Center for Materials Nanoarchitectonics, National Institute for Materials Science, 1-1 Namiki, Tsukuba 305-0044, Japan*
*[6]Research Center for Electronic and Optical Materials, National Institute for Materials Science, 1-1 Namiki, Tsukuba 305-0044, Japan*
**Corresponding Author*

1. **Figure S1: Near-field approach curve**

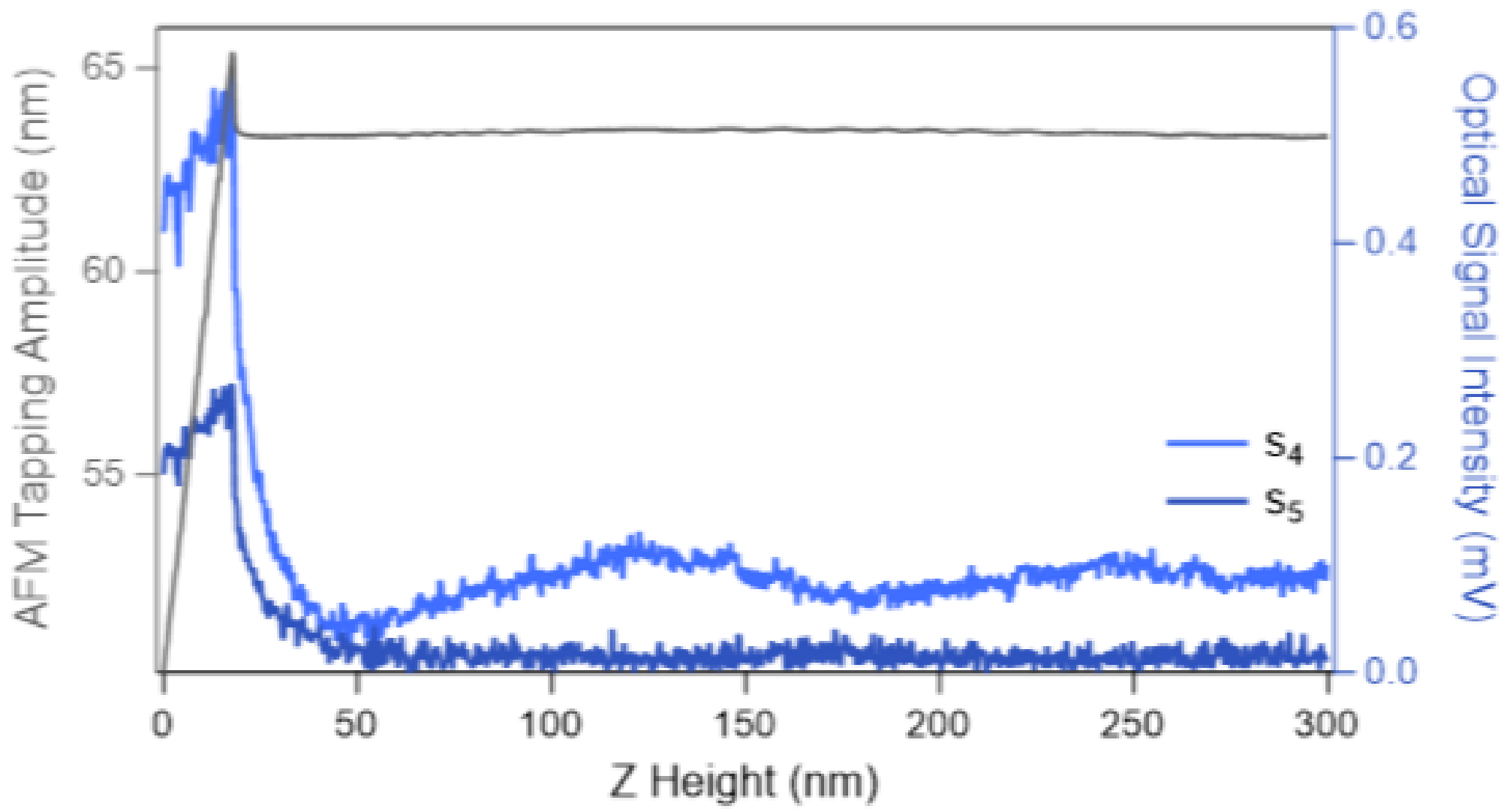

**Figure S1:** Near field approach curve. Representative data taken on $WSe_2$ at 1.73 eV and 9 K. The $s_4$ and $s_5$ amplitude signals decay exponentially as the illuminated AFM tip retracts 300 nm. The exponential decay begins where the tip-sample interaction terminates – shown by the point at which the tapping amplitude becomes constant, at a Z-height retraction of ~25 nm along the x-axis.

## 2. Figure S2: Nano-indentation AFM

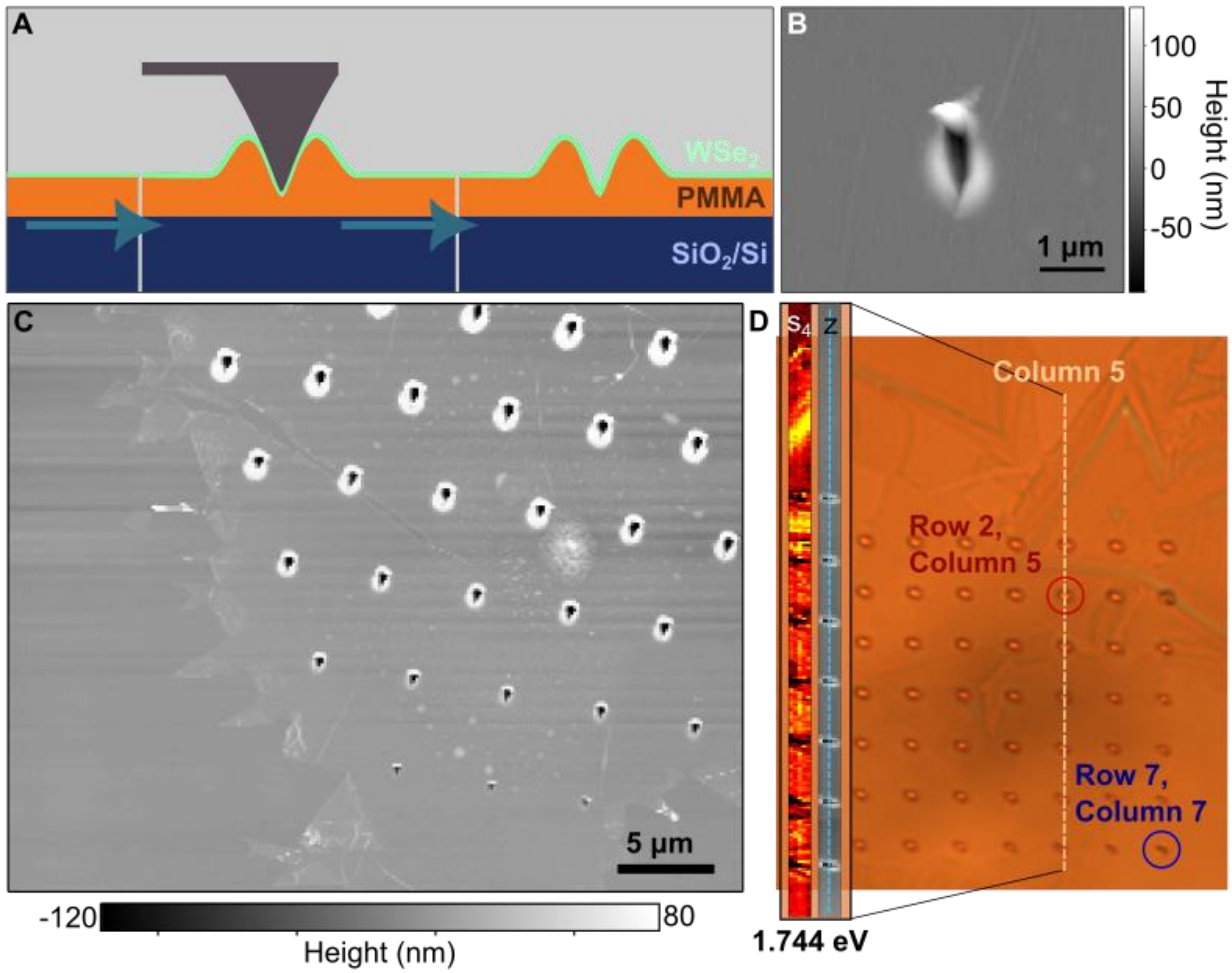

**Figure S2:** Nanoindentation technique **A** Schematic of AFM indentation procedure. **B** AFM topography of a representative indentation region, used to model strain levels in Figure S6. **C** By adjusting the Z-depth parameter, arrays of indents with varying depths and strain profiles can be produced as shown in C from the AFM image of Sample 2. **D** Schematic of the line-scan measurements in Figure 2. The optical image of the sample is shown, with the corresponding topography along Column 5 and near-field amplitude at 1.744 eV. A line scan along the blue dashed line in the Z image is repeated as a function of energy to create the line scan plots in Figure 2 and similarly in Figure 5.

## 3. Figure S3: Bare PMMA near-field response

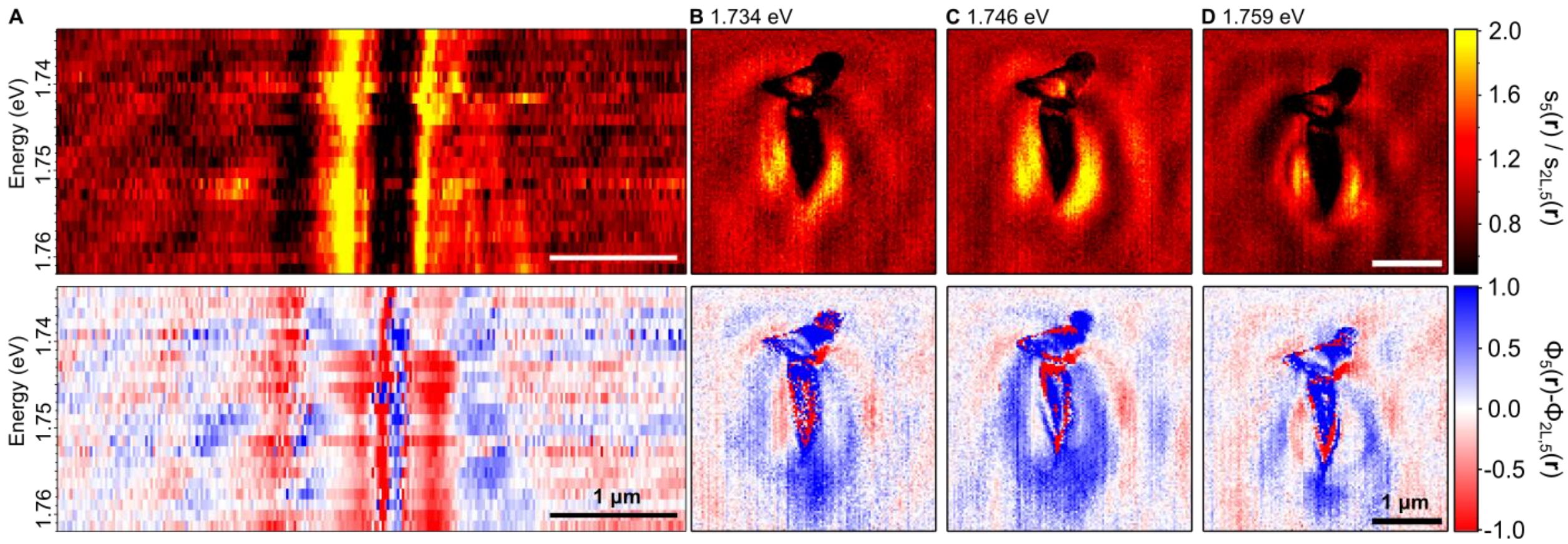

**Figure S3:** PMMA indent near-field response. To confirm the observed energy shifts are from the $WSe_2$ excitonic response and not the unusual deformation profile in the underlying PMMA, we repeat the measurements from Figure 2 and Figure 3 on bare PMMA. Figure S3 shows no near-field response in the measured spectral range beyond 1 micron from the indent center. **A** Variation of the near-field amplitude $s_5$ (top) and phase $\phi_5$ (bottom) signal across the bare PMMA indent scan shown in Figure 1B as a function of excitation energy (left axis). **B-D** (top) Normalized amplitude $s_5$ image of the sample taken at excitation energies 1.734, 1.746, and 1.759 eV. (bottom) Corresponding normalized phase $\phi_5$images measured at the same energy shown in the above amplitude scans. Scale bars are 1 micron.

## 4. Finite Dipole Modeling

We extracted the dielectric response using Finite Dipole Model (FDM) fitting, implemented via the open-source Python package snompy[1]. To capture the layered geometry of the sample, the sample was modeled as a multilayer stack following the transfer matrix method[2,3]. From top to bottom, the layers and their thicknesses were: air (infinite environment), $WSe_2$ (0.7 nm), and PMMA (infinite substrate) for the nano-indentation sample and air (infinite environment), hBN (20 nm), $WSe_2$ (0.7 nm), and $SiO_2$ for the pillar sample. The dielectric functions of hBN, PMMA, and $SiO_2$ in the visible range were taken as 3.8, 2.2, and 2.1, respectively. The ellipsoidal tip model had a length of 350 nm and a radius of curvature of 25 nm to match the geometry of the AFM probe used in this work. The complex dielectric response of $WSe_2$ was described by a Lorentz oscillator model, determined through a recursive fitting procedure that combined our near-field spectra with the FDM simulation results. From the complex Lorentz oscillator equation,

$$\varepsilon(\omega) = \varepsilon_\infty - \frac{\hbar c}{\omega_0 d} \frac{\gamma_r}{(\omega_0 - \omega) + i\left(\frac{\gamma_{nr}}{2} + \gamma_d\right)}$$

the high frequency permittivity of the sample ($\varepsilon_\infty$), the center energy ($\omega_0$) of the oscillator shown in Figure 3D, and the radiative ($\gamma_r$), non-radiative and dephasing decay rates $\left(\frac{\gamma_{nr}}{2} + \gamma_d\right)$ were the input parameters of the recursive fit. These values were varied to minimize the difference between the calculated near field response and measured data.

## 5. Figure S4: Far-field reflectivity

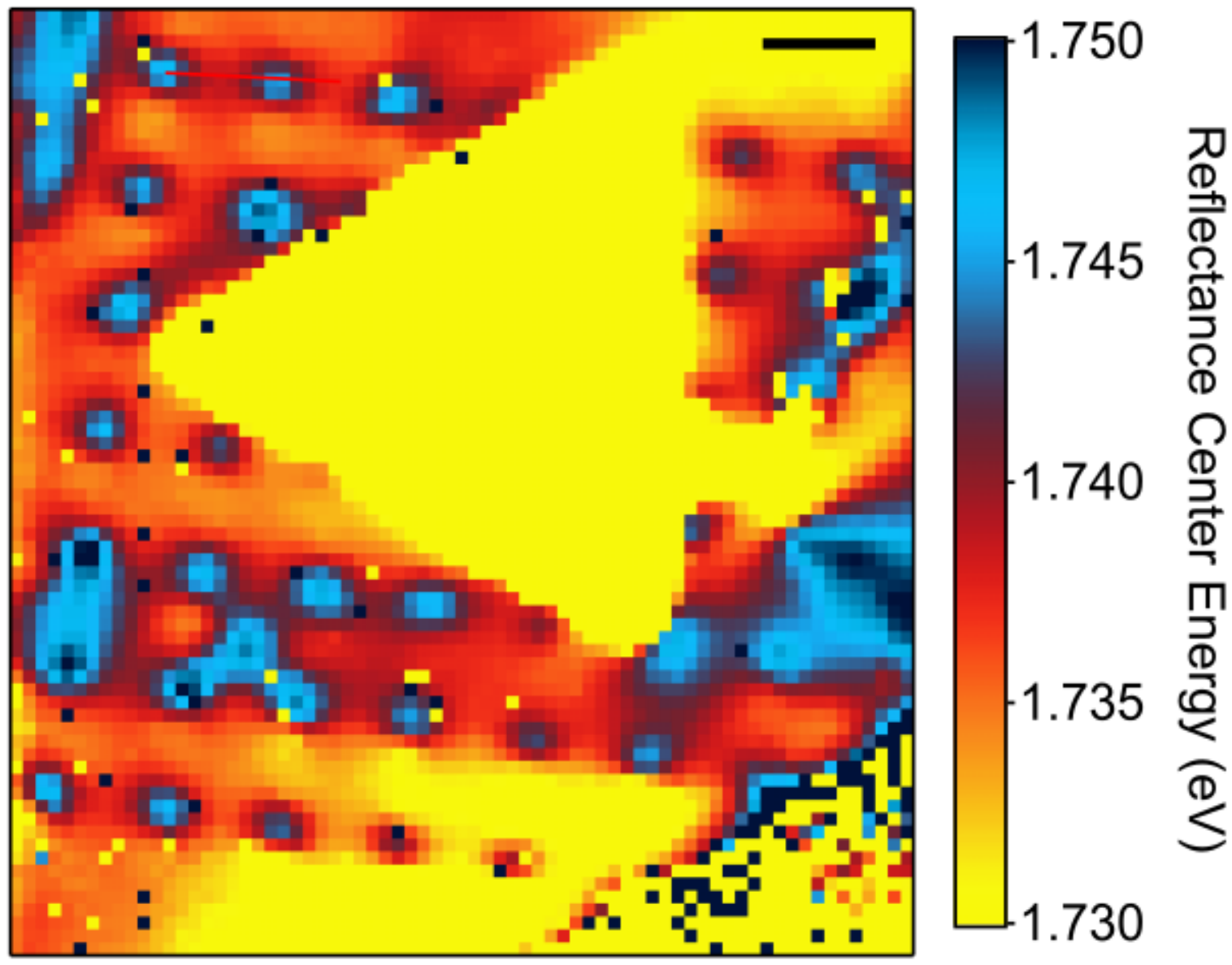

**Figure S4:** Representative far-field reflectivity showing $WSe_2$ neutral exciton resonance center energy across an AFM indented sample. Scale bar is 5 microns. Far-field reflectance spectra were acquired at 25 K using a commercial Horiba LabRam Evolution confocal spectrometer integrated with a Montana Instruments closed-cycle optical cryostat. Illumination was provided by a laser-driven light source (Energetiq Technology, EQ-99X-S LDLS). The white light was focused onto the sample surface using an Olympus 50× objective (SLMPLN50x), resulting in a spot size of approximately 2 µm. To map the sample, high-speed galvanometer (galvo) mirrors were used to scan the focused spot across a 30 µm×30 µm region with a 0.5 µm step size. To minimize optical contributions from the substrate, light source, and system components, the reflectance contrast ($RC = (R_{off} - R_{on})/R_{off}$) center energy is plotted. Here, $R_{on}$ represents the $WSe_2$ region of interest and $R_{off}$ is the reference spectrum from the bare $SiO_2/Si$ substrate.

## 6. Figure S5: Unstrained $WSe_2$ A exciton resonance

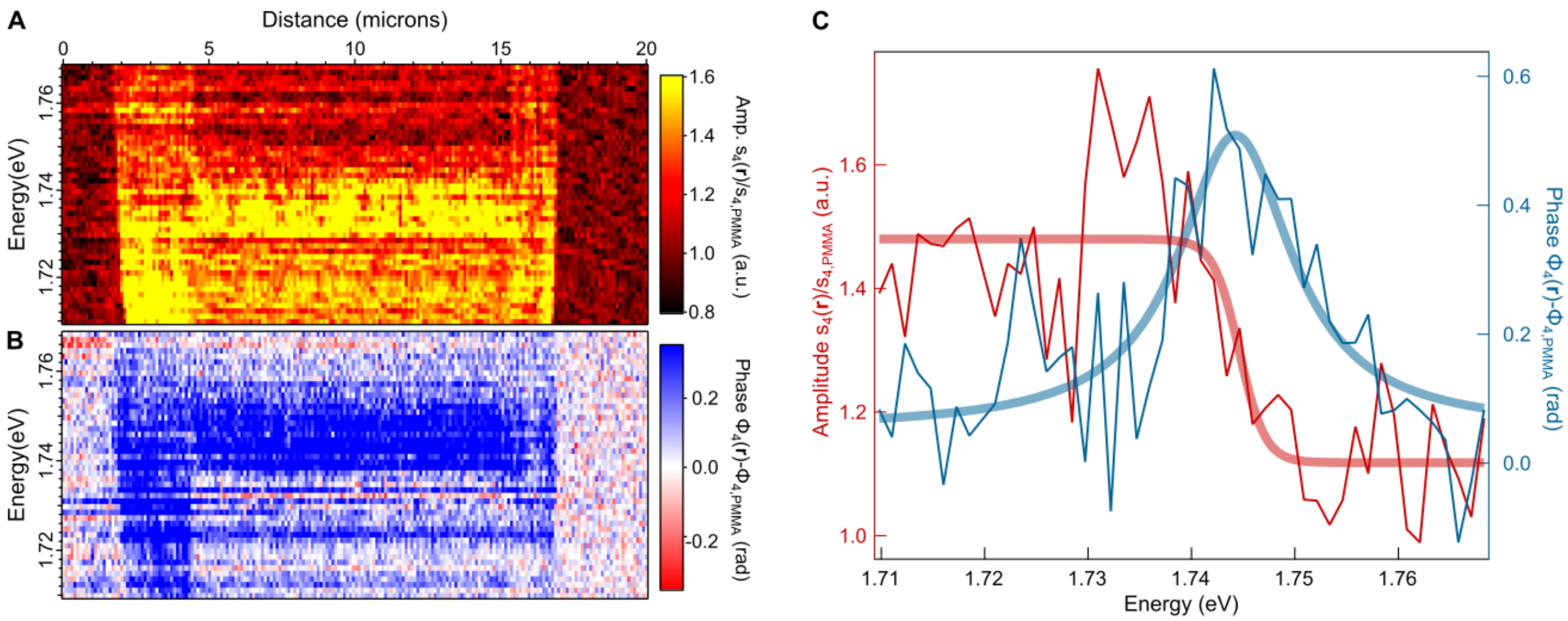

**Figure S5:** Unstrained region of $WSe_2$ on PMMA used to compare the shifted A exciton resonance in the nano-indented regions. **A.** Variation of the near-field amplitude $s_4$ across a region of monolayer $WSe_2$ located from 5 to 17 microns in the repeated line scan as a function of excitation energy. The darker regions from 0-1.5 and 17 to 20 microns are PMMA used to normalize the TMD energy dependent response. Region 1.5 to 5 microns is a fold in the device. **B** corresponding phase $\phi_4$ signal as a function of line scan position and excitation energy. **C** Vertical line profiles across the resonance mapped in **A** and **B**, showing an A exciton resonance at 1.745 eV in good agreement with the unstrained assignment in Figure 2.

7. **Strain Modelling**

a. Calculations

Following previous works[4], we modeled the strained indentation as a TMD monolayer draped over the deformed PMMA. This deformation is given by the out of plane displacement, $h(x, y)$, or the measured height profile (Nano-indent Figure 1B, row 2 column 5 indent, AFM: Figure S2B and pillar AFM: Figure 5B). The monolayer is treated as a plate with infinitesimal thickness. In this 2D limit, the strain tensor $\epsilon(\boldsymbol{r_{xy}})$ is given by

$$\epsilon_{ij} = \frac{1}{2}\left(\partial_i u_j + \partial_j u_i + \partial_i h\, \partial_j h\right) \tag{1}$$

where $u_x$ and $u_y$ are the in-plane displacements along x and y and can be solved for with known out-of-plane displacement $h(x, y)$. The in-plane force equilibrium condition ensures the material equilibrium is maintained at position $(r)$ and can be written in terms of the stress tensor, σ:

$$\partial_x \sigma_{xx} + \partial_y \sigma_{xy} = 0, \tag{2A}$$

$$\partial_x \sigma_{xy} + \partial_y \sigma_{yy} = 0 \tag{2B}$$

The strain and stress tensors are related using Hooke's law as

$$\begin{pmatrix} \sigma_{xx} \\ \sigma_{yy} \end{pmatrix} = \frac{E}{1-\nu^2} \begin{pmatrix} 1 & \nu \\ \nu & 1 \end{pmatrix} \begin{pmatrix} \epsilon_{xx} \\ \epsilon_{yy} \end{pmatrix}, \sigma_{xy} = \frac{E}{1+\nu}\epsilon_{xy} \tag{3}$$

where E is the materials Young's modulus and ν is the Poisson's ratio. Inserting Eq.1 and 3 into Eq. 2 yields a coupled PDE for $u_x$ and $u_y$ which once solved can be used in Eq. 1 to calculate the total strain $\epsilon_{tot} = \epsilon_{xx} + \epsilon_{yy}$.

$$\partial_{xx}^2 u_x + \frac{1-\nu}{2}\partial_{yy}^2 u_x + \frac{1+\nu}{2}\partial_{xy}^2 u_y + \partial_x h\left(\partial_{xx}^2 h + \frac{1-\nu}{2}\partial_{yy}^2 h\right) + \frac{1+\nu}{2}\partial_y h\, \partial_{xy}^2 h = 0,$$

$$\partial_{yy}^2 u_y + \frac{1-\nu}{2}\partial_{xx}^2 u_y + \frac{1+\nu}{2}\partial_{xy}^2 u_x + \partial_y h\left(\partial_{yy}^2 h + \frac{1-\nu}{2}\partial_{xx}^2 h\right) + \frac{1+\nu}{2}\partial_x h\, \partial_{xy}^2 h = 0 \tag{4}$$

Taking $\nu = 0.19$ [5] Eq. 4 can be numerically solved for $u_x$ and $u_y$ and the total strain is then calculated from the now known in-plane and out-of-plane displacements, $u_x$, $u_y$, and $h(x, y)$.

The total strain map measured from the AFM deformation height profile (Supplementary Figure S6) agrees with previous work, showing several percent strain roughly 0.5 microns from the indent center. However, in this configuration, the strain within the indent center surpasses 10 percent, beyond the tearing strain of TMD monolayers experimentally measured to occur beginning at and above 6 percent strain[6], while the strain at the top of the indent, where the topography reaches its highest value, reaches 40 percent. The A exciton energy corresponding to modelled strain levels can be estimated as -63 meV/percent for tensile strain from previous measurements of strained monolayer TMDs on a polypropylene substrate[7]. Such large strain would shift the A exciton energy by over 400 meV near the indentation, far beyond the 35 meV window scanned in our near field measurements around the A exciton resonance. However, Figure 2 shows clear spatial variation of the A exciton resonance around the indentations within this small energy range. An explanation for the smaller energy shifts directly measured from near-field probing is that the TMD monolayer is not stretched over the deformed indent, but rather tore, which would significantly reduce the overall strain. Tears can be seen in high resolution AFM images (Figure S7) around indentations, before the extreme slope masks potential additional tears within the indent crater and raised rim. Recent SEM imaging has similarly determined nano-indentation can tear TMD monolayers[8].

### b. Figure S6: Strain mapping

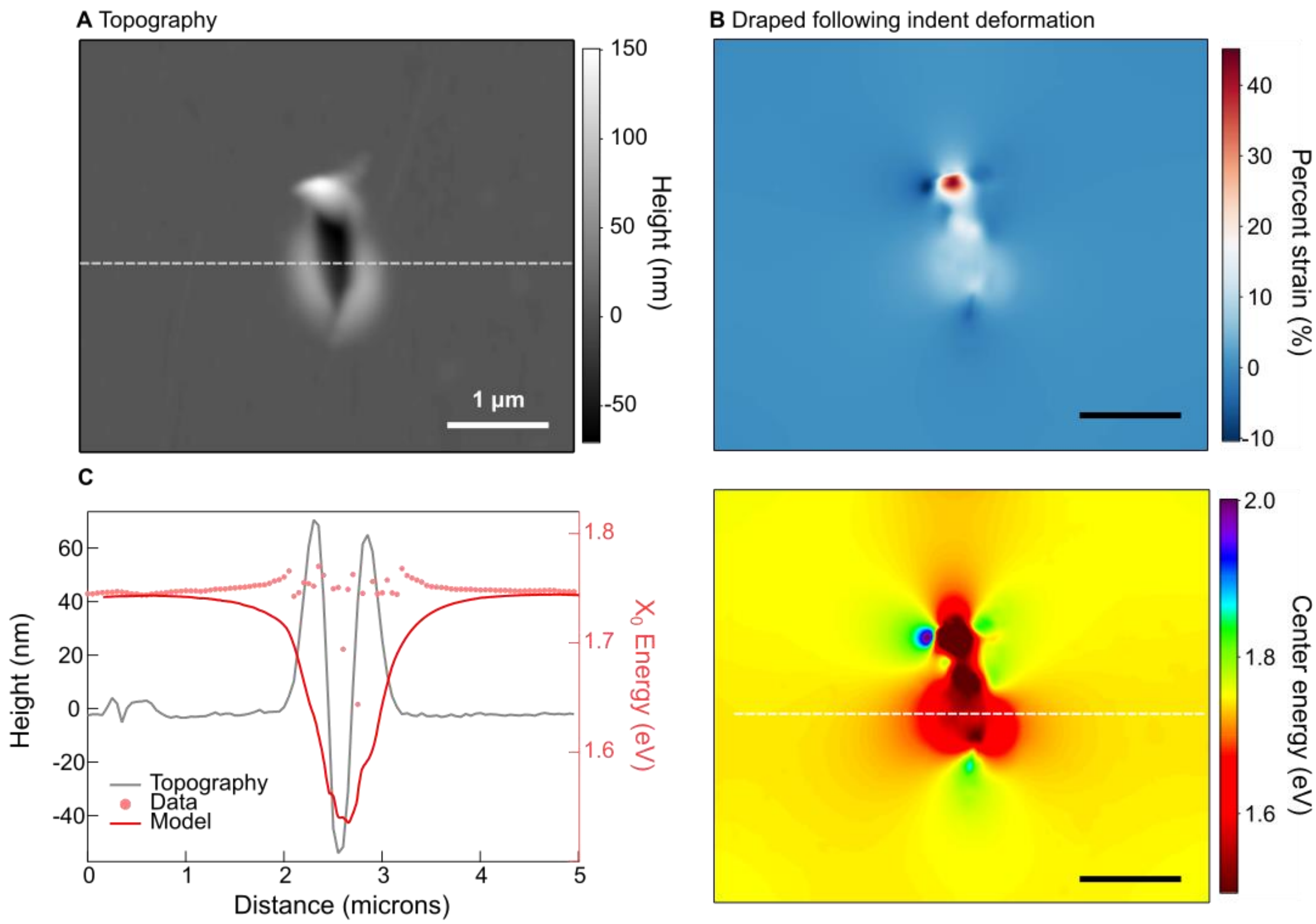

**Figure S6:** Total strain mapping, assuming untorn TMD **A** Sample topography **B** Top: Total strain percentage across the region mapped in **A**. Bottom: Calculated A exciton energy assuming strain percentages shown above. **C** Line cuts through the indent topography overlaid with the modelled (line) and measured (markers)A exciton energy images, corresponding with the line scan orientation shown in Figure 2. Shaded red box shows the energy range measured in Figure 2 and 3 where the A exciton energy was observed to shift roughly 15 meV around the indent.

c. **Figure S7: AFM of material tearing**

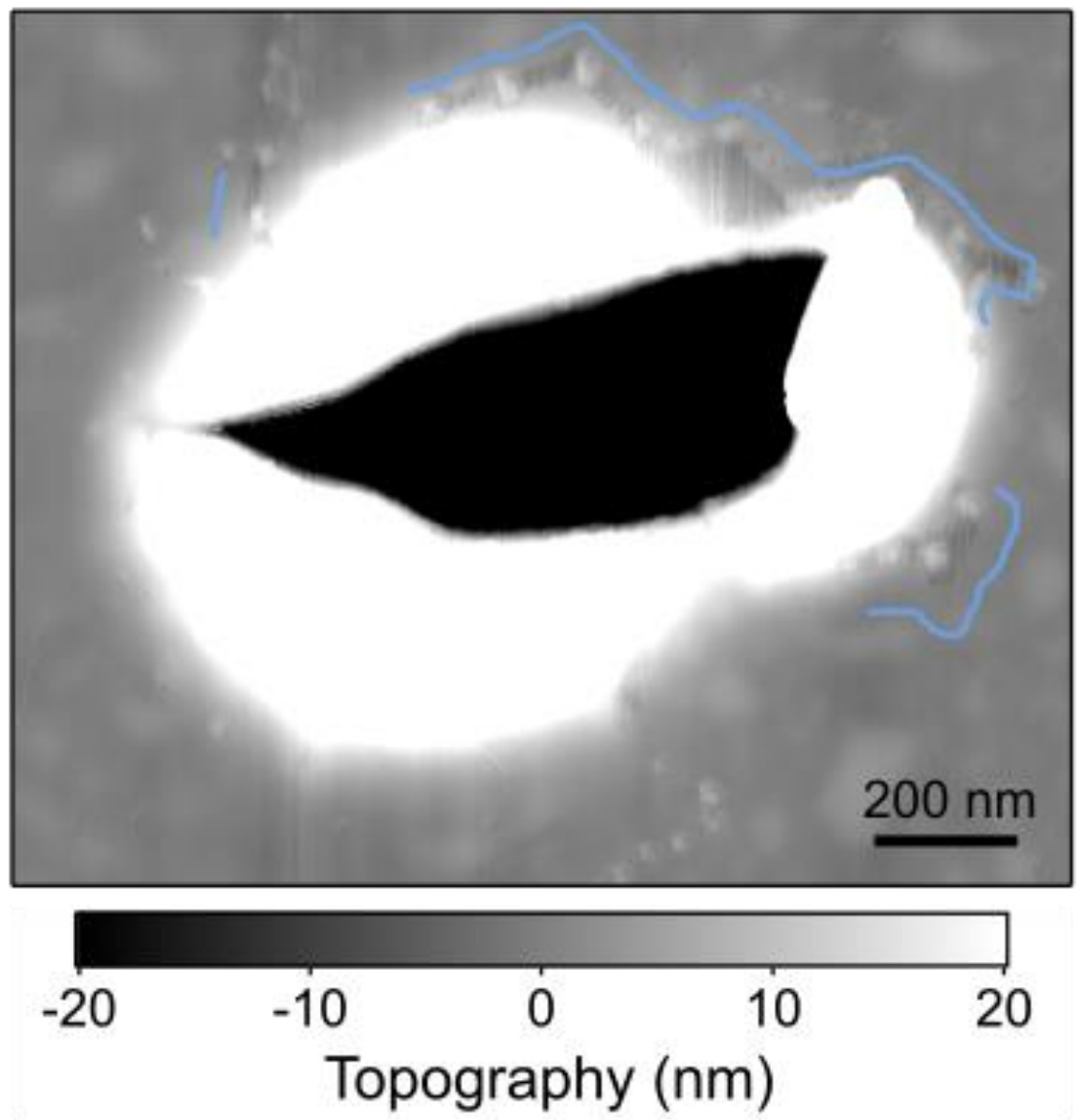

**Figure S7:** Representative indent showing regions of tears in the TMD monolayer around the nanoindentation site. Outer edges of the observed tears are outlined in blue. Tears are visible by the slightly darker (lower) regions in the topography near the high (white) indent rim, saturated by the color scale.

## 8. Figure S8: Optical image of Sample 2

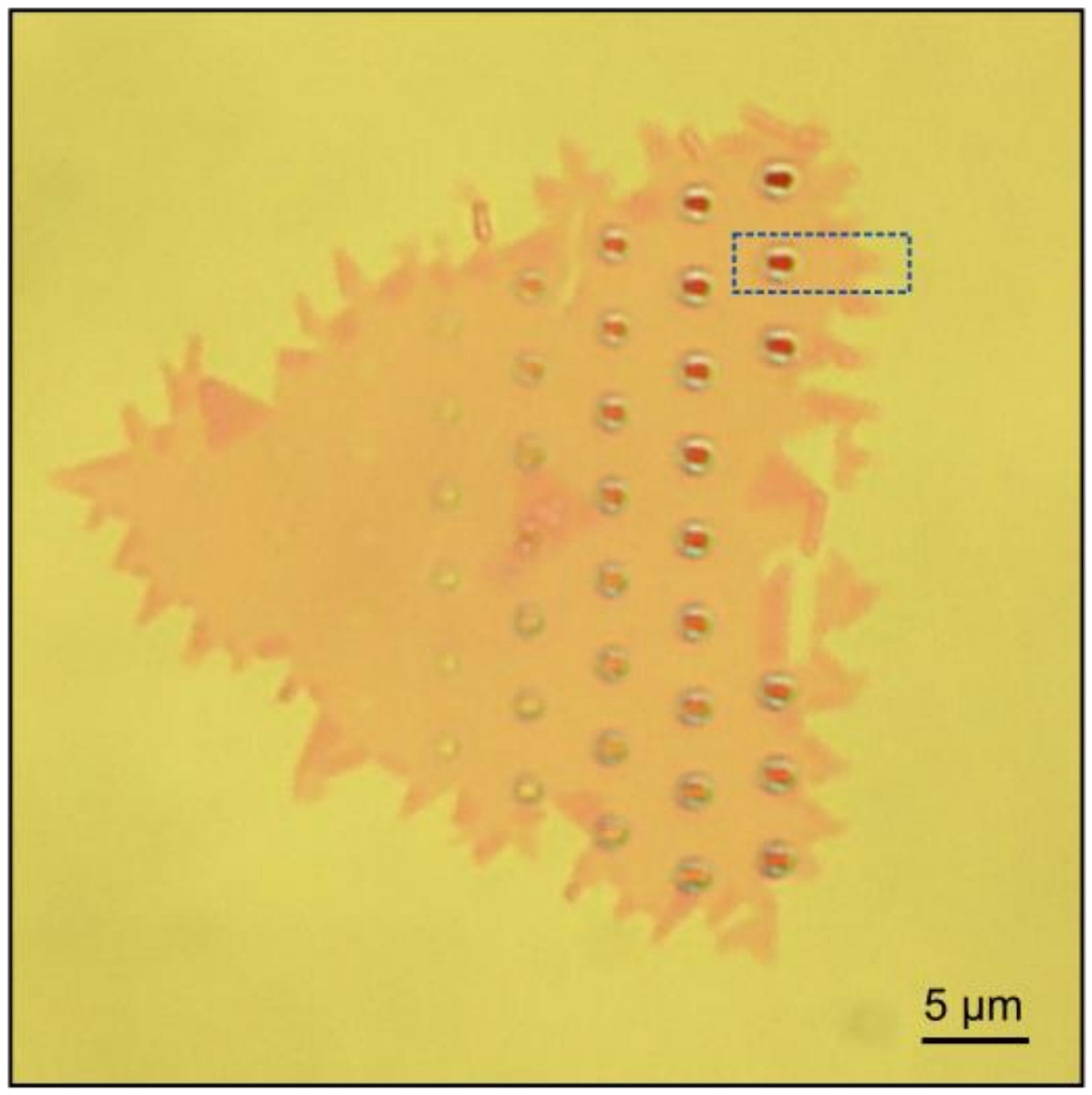

**Figure S8:** Optical image of Sample 2 measured in Figure 3, with indent area measured shown by blue dashed rectangle. Scale bar: 5 microns. This device was fabricated with varying indentation depths, as shown in the corresponding AFM image in Figure S2.

## 9. Figure S9: Additional pillar line scan comparison

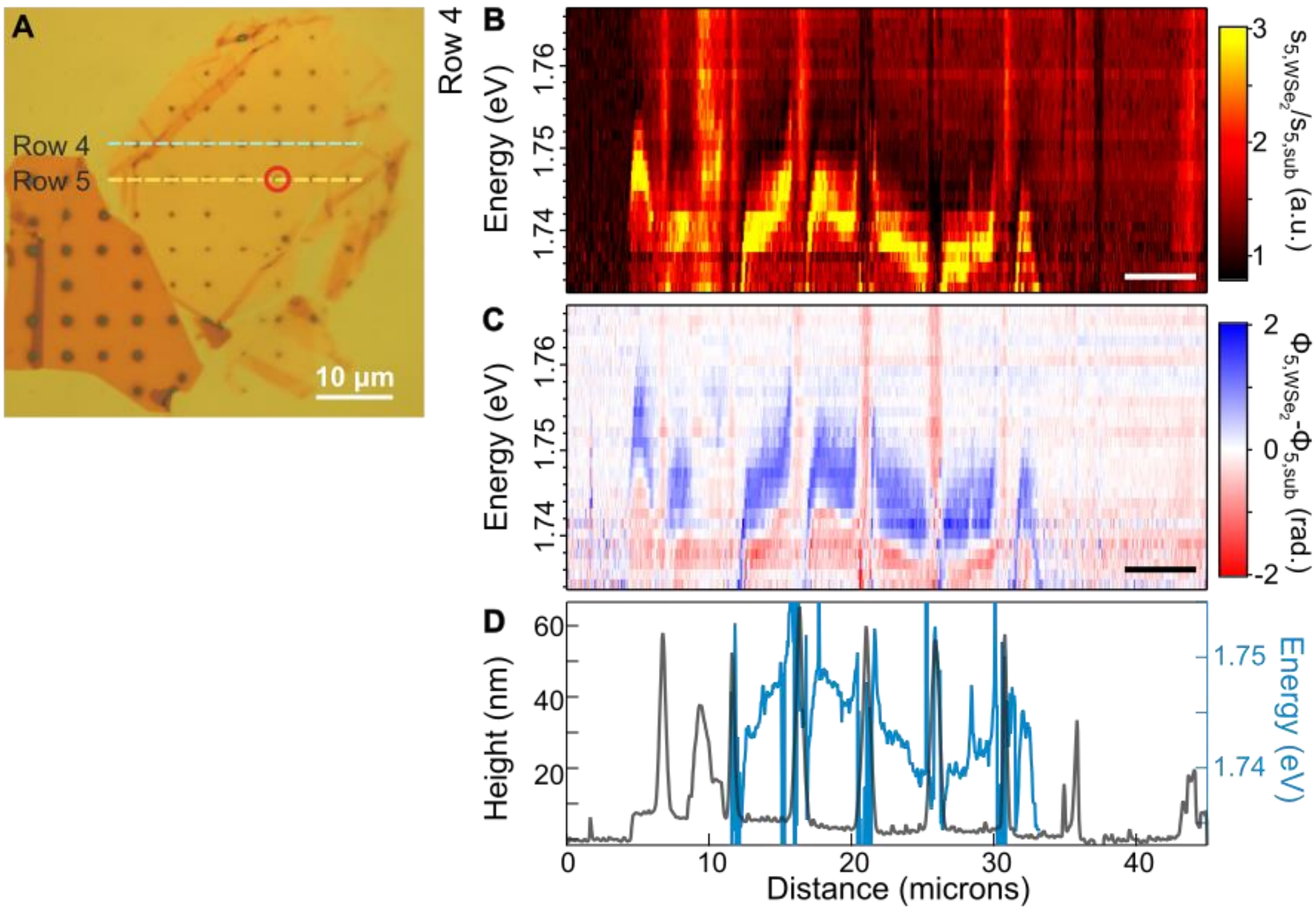

**Figure S9:** Measurements and analysis from Figure 5C-E were repeated along a second row of the pillar strained device and depict a comparable energy landscape. **A** Optical image of pillar strained device showing deposited hBN capped $WSe_2$ on a grid of 60 nm tall cylindrical pillars with 5 micron spacing. **B** Variation of the near-field amplitude $s_5$ across Row 4 shown in **A** as a function of excitation energy (left axis). **C** corresponding phase $\phi_5$ signal. Scale bars are 5 microns. **D** Sample height profile (gray, left axis) along the scanned line in **B,C** with finite dipole model (FDM) extracted center energy of the A exciton resonance extracted from a vertical line cut at each pixel (blue, right axis). Extracted center energy values shown for the region of clearly resolvable resonance behavior in **B, C**.

### 10. Figure S10: Nanoindentation $g^{(2)}$ measurement

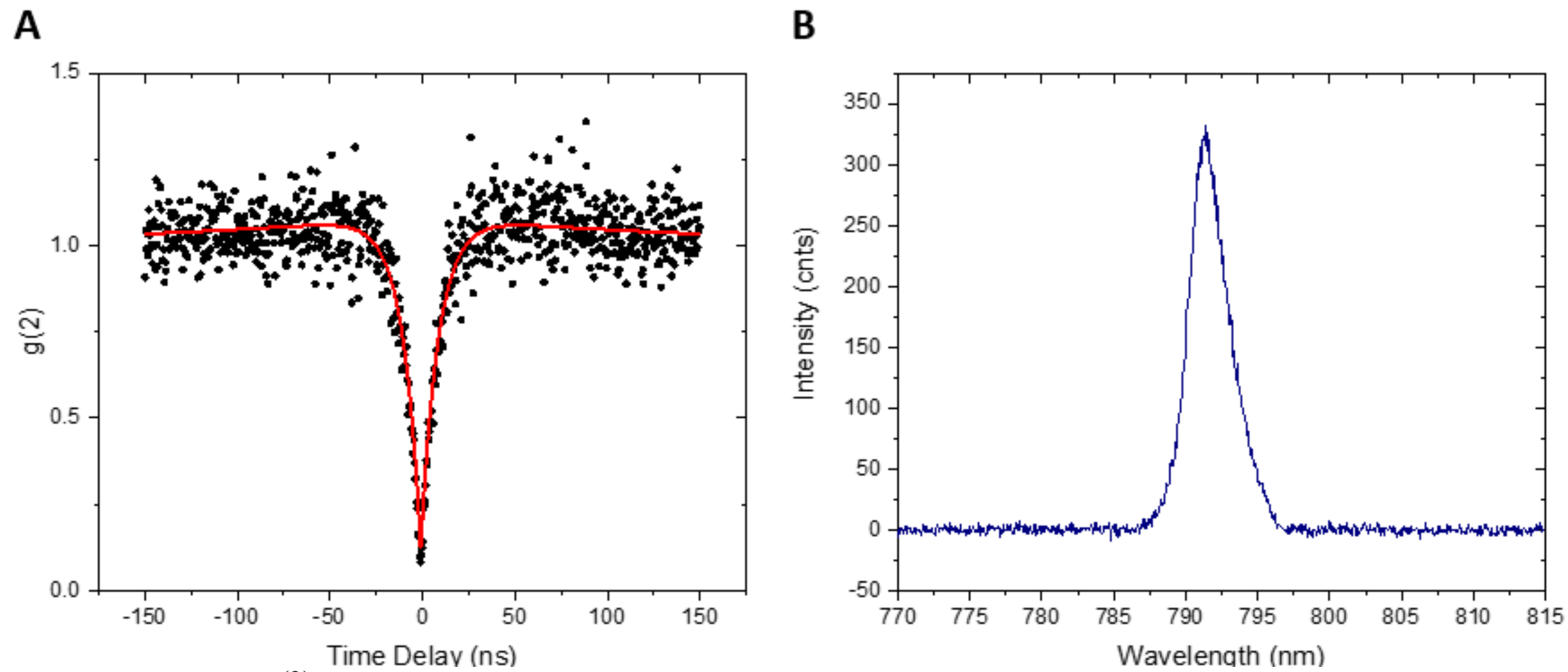

**Figure S10: A** $g^{(2)}$ measurement taken on the indent marked by the blue circle in Figure 1, as well as shown in the topography in Figure S6A. $g^{(2)}(0)<0.1$ shows high-purity emission. **B** Corresponding PL spectra.

### 11. Figure S11: Pillar $g^{(2)}$ measurement

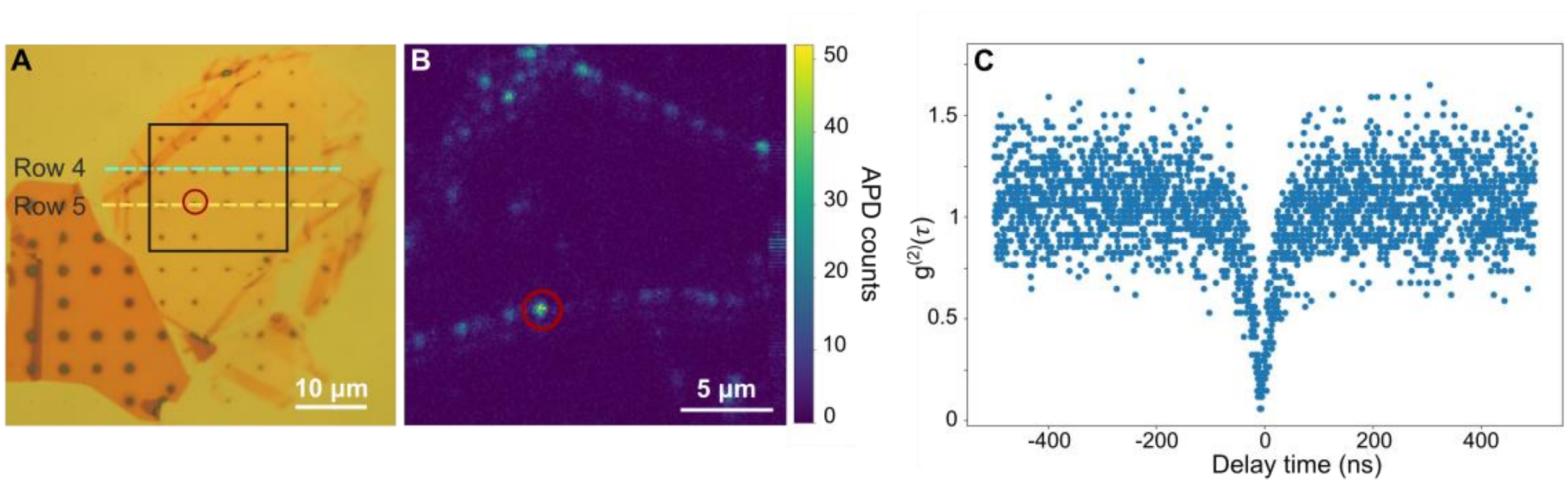

**Figure S11: A** Pillar device from Figure 5 and Figure S9. **B** Corresponding emission map showing avalanche photodiode (APD) counts. **C** $g^{(2)}$ measurement taken on the pillar marked by the red circle in A and B.